\documentclass[preprint,english,showpacs,11pt,floatfix]{revtex4}

\usepackage{babel,amsmath,amssymb,dcolumn}
\usepackage[dvips]{graphics}
\usepackage{hyperref}

\usepackage{amsfonts}
\usepackage{amssymb}
\usepackage[dvips]{graphicx}
\usepackage{latexsym}

\usepackage{dcolumn}
\usepackage{bm}

\newcommand{\be}{\begin{equation}}
\newcommand{\ee}{\end{equation}}

\begin{document}

\title{Constraints on the interacting
holographic dark energy model}

\author{Bin Wang}
\email{wangb@fudan.edu.cn} \affiliation{Department of Physics,
Fudan University, 200433 Shanghai}

\author{Chi-Yong Lin}
\email{lcyong@mail.ndhu.edu.tw}
\affiliation{Department of Physics, National Dong Hwa University,
Shoufeng, 974 Hualien}

\author{Elcio Abdalla}
\email{eabdalla@fma.if.usp.br} \affiliation{Instituto de Fisica,
Universidade de Sao Paulo, C.P.66.318, CEP 05315-970, Sao Paulo}

\begin{abstract}
We examined the interacting holographic dark energy model in a universe
with spatial curvature. Using the near-flatness condition and requiring
that the universe is experiencing an accelerated expansion, we have
constrained the parameter space of the model and found that the model can
accommodate a transition of the dark energy from $\omega_D>-1$ to
$\omega_D<-1$.

\end{abstract}


\maketitle

Numerous observational results indicate that our universe is undergoing an
accelerated expansion driven by a yet unknown dark energy (DE)
\cite{1}. While the leading interpretation of such a DE is a cosmological
constant whose equation of state (EOS) is $\frac p \rho
\equiv\omega_D=-1$, other conjectures relate  DE to a socalled
quintessence, with $\omega_D>-1$, or yet to an exotic field with
$\omega_D<-1$\cite{2}. An extensive analysis finds that the current data
favors DE models with EOS in the vicinity of $\omega_D=-1$ \cite{3},
straddling the cosmological constant boundary. Further works on the data
analysis can be traced in \cite{pad} Recently, the analysis of
the type Ia supernova data indicates that the time varying DE gives a
better fit than a cosmological constant \cite{4}, which mildly favor the
evolution of the DE EOS from $w_D>-1$ to $w_D<-1$ at a recent
stage. Theoretical attempts towards understanding of the $\omega_D$
crossing $-1$ phenomenon have taken place \cite{5}.

In a recent work we proposed a holographic DE model\footnote{For most
discussions on holographic DE independently evolving from other matter
fields see \cite{6,7,8,8b,9}} with interaction with matter fields
to explain the above transition of the DE \cite{10}. Given the unknown
nature of both DE and dark matter (DM), which are two major contents of
the universe, one might argue that an entirely independent behavior of DE
is very special \cite{11}. Studies on the interaction between DE and DM
have been carried out \cite{11,12,13}. It was argued that the interaction
will influence the perturbation dynamics and could be observable through
the lowest multipoles of CMB spectrum \cite{12}. Investigation of the
interaction between DE and DM in the holographic DE model has been done by
using the Hubble scale as IR cutoff to explain the acceleration of our
universe \cite{14}. In \cite{10}, we extended  the inclusion of
interaction between DE and DM into the holographic DE model with the
future event horizon as an IR cutoff. As a result, we found that our
model, with the interaction between DE and DM, can give an early
deceleration and late a time acceleration. In addition, the appropriate
coupling between DE and DM accommodates the transition of the DE equation
of state from $w_D>-1$ to $w_D<-1$. This property could serve as an
observable feature of the interaction between DE and DM, in addition to
its influence on the small $l$ CMB spectrum argued in \cite{11}.

In this paper we would like to extend our previous discussion \cite{10} to
a universe with spatial curvature (see also \cite{franca}). 
The tendency of preferring a closed
universe appeared in a suite of CMB experiments \cite{15}. The improved
precision from WMAP provides further confidence, showing that a closed
universe with positively curved space is marginally preferred
\cite{16}. In addition to CMB, recently the spatial geometry of the
universe was probed by supernova measurements of the cubic correction to
the luminosity distance \cite{17}, where a closed universe is also
marginally favored. At present, the ratio of the sum of the densities of
all forms of matter energy in the universe to the critical density
required for spatial flatness is $\Omega_{T,0}=1.02\pm 0.02$. We will use
this ``near flatness" property together with the transition of the EOS of
DE to constrain our model parameters.

The total energy density is $\rho=\rho_m+\rho_D$, where $\rho_m$ is the
energy density of matter and $\rho_D$ is the energy density of the DE. The
total energy density satisfies a conservation law. However since we
consider the interaction between DE and DM, $\rho_m, \rho_D$ do not
conserve separately. They must rather enter the energy balances
\cite{14}\cite{10}
\begin{eqnarray}
\dot{\rho}_m+3H\rho_m&=&Q\quad ,\label{1}\\
\dot{\rho}_D+3H(1+w_D)\rho_D&=&-Q\quad ,\label{2}
\end{eqnarray}
where $w_D$ is the equation of state of DE, $Q$ denotes the interaction
term and can be taken as $Q=3b^2 H\rho$ with $b^2$ the coupling constant
\cite{15}. This expression for the interaction term was first introduced
in the study of the suitable coupling between a quintessence scalar
field and a pressure-less cold dark matter field \cite{11}.
The choice of the interaction between both components
was meant to get a scaling solution to the coincidence
problem such that the universe approaches a stationary
stage in which the ratio of dark energy and dark matter
becomes a constant. In the context of holographic
DE model, this form of interaction was derived from
the choice of Hubble scale as the IR cutoff \cite{14}.

From (\ref{1}), (\ref{2}) and the Friedmann equation with positive
curvature  $\Omega_m+\Omega_D=1+\Omega_k$, where $\Omega_m=\rho_m/(3H^2)$,
$\Omega_D=\rho_D/(3H^2)$ and $\Omega_k=1/(aH)^2$, we get the equation of
state of DE,
\be
\omega_D=\frac{\Omega_k'}{3(1+\Omega_k-\Omega_D)}-\frac{(1+\Omega_k)
\Omega_D'}{3\Omega_D(1+\Omega_k-\Omega_D)}-\frac{b^2(1+\Omega_k)^2}
{\Omega_D(1+\Omega_k-\Omega_D)}\quad ,\label{3}
\ee
where the prime denotes the derivative with respect to $x=\ln a$.

The holographic DE in the closed universe is expressed as
$\rho_D=3c^2 L^{-2}$  where $c^2$ is a constant \cite{7} and with the
event horizon as the IR cutoff. Here $L=ap(t)$, where $p(t)$ is defined by
$\int^{p(t)}_0 \frac{dp}{\sqrt{1-k p^2}}=\int^{\infty}_t\frac{dt}{a}=R_h/a$ or
$p(t)=\sin y/\sqrt{k}$, where $y=\sqrt{k} R_h/a$. $R_h$ is the radial size
of the event horizon and $L$ is the radius of the event horizon measured on
the sphere of the horizon.

The DE density can also be written as $\rho_D=3c^2 L^{-2}=\Omega_D
3H^2$. Thus we have $L=\frac{c}{H\sqrt{\Omega_D}}=a\sin y/\sqrt{k}$.
Taking the derivative with respect to $t$ on both sides of such an
equation, we have
\be
-\cos
y=-\frac{c}{\sqrt{\Omega_D}}(1+\dot{H}/H^2)-
\frac{c\dot{\Omega_D}}{2\Omega_D^{3/2}H} \label{4}
\ee

Deriving now the Friedmann equation with respect to $t$ and using
eqs(\ref{1},\ref{2}), we have
\be
\dot{H}/H^2=\frac{-3\Omega_D(1+r+\omega_D)/2-
\dot{\Omega_k}/(2H)}{1+\Omega_k}\quad ,\label{5}
\ee
where $r$ is the ratio of the energy densities,
$r=\rho_m/\rho_D=(1+\Omega_k-\Omega_D)/\Omega_D$. Substituting (\ref{5})
into (\ref{4}) and considering the expression of $\omega_D$ we got before,
we obtain the evolution behavior of the dark energy,
\be
\frac{\Omega_D'}{\Omega_D^2}=\frac{1+\Omega_k-\Omega_D}{1+\Omega_k}
[\frac{2\cos y}{c\sqrt{\Omega_D}}+\frac{1}{\Omega_D}+\frac{\Omega_k'}
{\Omega_D(1+\Omega_k-\Omega_D)}-\frac{3b^2(1+\Omega_k)}
{\Omega_D(1+\Omega_k-\Omega_D)}]\quad .\label{6}
\ee
Neglecting the interaction between DE and DM, namely $b^2=0$, this
result leads to (31) in \cite{7} if we substitute
$\Omega_k=aq\Omega_m$ where $q=\Omega_{k0}/\Omega_{m0}$. If we keep
$b^2$ but neglect the curvature of the universe, this expression
returns to (5) of reference \cite{10}.

With the expression of $\Omega_D'/\Omega_D^2$, we can rewrite (\ref{3}) in
the form
\be
\omega_D=-1/3-2\sqrt{\Omega_D}\cos y/(3c)-b^2(1+\Omega_k)
(1-\Omega_D)/[\Omega_D(1+\Omega_k-\Omega_D)]\quad ,\label{7}
\ee
where $\cos y=\sqrt{1-c^2\Omega_k/\Omega_D}$. In the derivation of
(\ref{7}), we have employed
\be
\Omega_k'=-2\Omega_k-2\Omega_k \times (H'/H)
\ee
and
\be
H'/H=-\frac{3\Omega_D(1+r+\omega_D)}{2(1+\Omega_k)}-\frac{\Omega_k'}
{2(1+\Omega_k)}=-3\Omega_D(1+r+\omega_D)/2+\Omega_k\quad .
\ee

With these equations at hand, we are in a position to study the evolution
behaviors of different forms of matter energy in the universe. We have two
parameters in the evolution equations, namely $b^2$ indicating the
coupling between the DE/DM and $c^2$ coming from the holography. Different
values of $b^2$ and $c^2$ influence a lot the evolution behavior of our
universe.

\begin{figure}[t]
\includegraphics[width=16cm,height=5cm,angle=0]{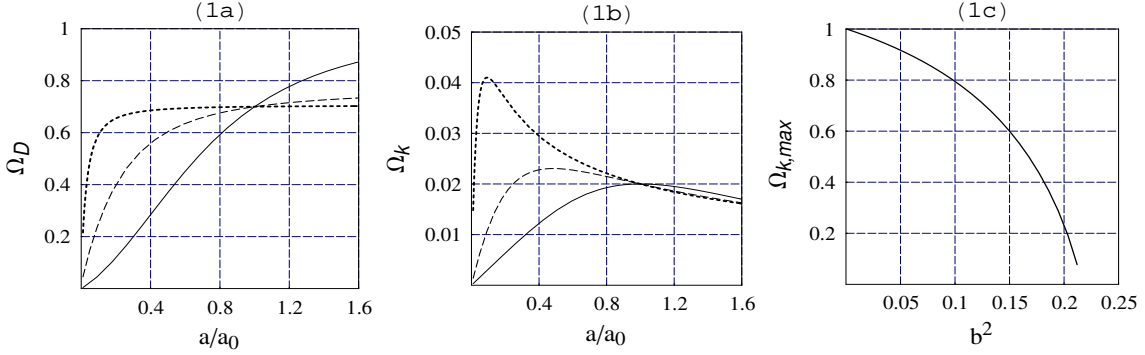}

\caption{In part $a$ we see the evolution of  $\Omega_D$ in terms of the
change of the coupling between DE and DM for a small fixed value of
$c$. In $b$ the evolution of $\Omega_k$ is shown in terms of the change of
$b^2$ for fixed small $c$. In $c$ the locations of peaks of $\Omega_k$
with the change of $b^2$ for the fixed small $c$ are pictured.} \label{f1}
\end{figure}
\begin{figure}[t]
\includegraphics[width=5.5cm,height=5.5cm,angle=0]{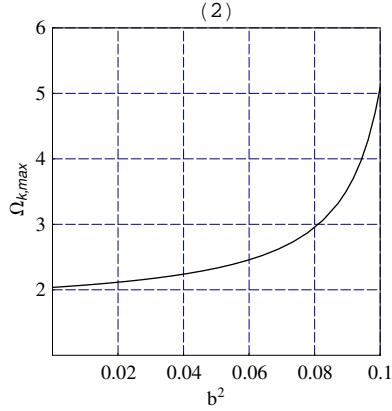}
\caption{Locations of peaks with the change of $b^2$ for a fixed big value of $c$.} \label{f2}
\end{figure}
\begin{figure}[t]
\includegraphics[width=16cm,height=5cm,angle=0]{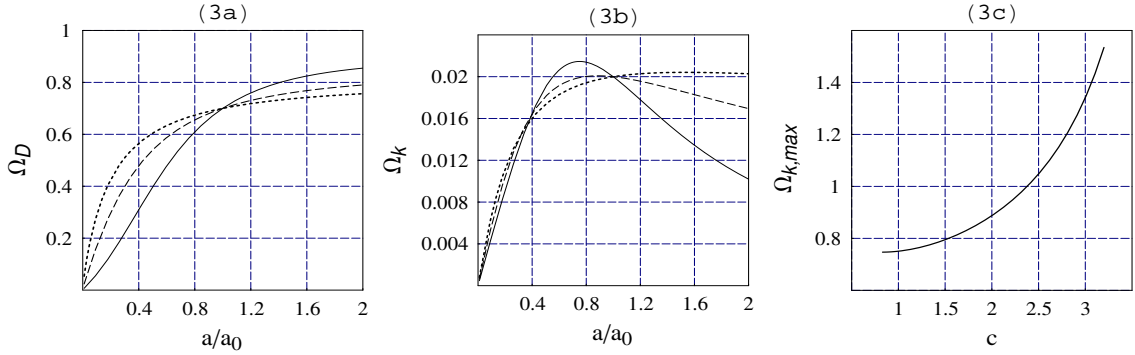}

\caption{In the part $a$ we show the evolution of $\Omega_D$ with the
change of $c$ for fixed small coupling between DE and DM. In $b$ we show
the evolution of $\Omega_k$ with the change of $c$ for fixed small value
of $b^2$. Part $c$ exhibits the locations of peaks with the change of
$c$. } \label{f1-2}
\end{figure}
\begin{figure}[t]
\includegraphics[width=16cm,height=5cm,angle=0]{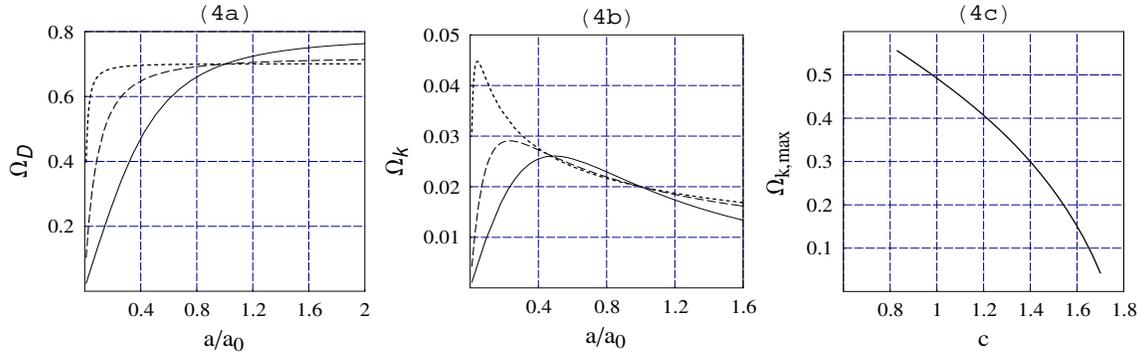}

\caption{Figure $4a$ shows the evolution behavior of $\Omega_D$ with the
change of $c$ for a large value of $b^2$. Figure $4b$, shows the behavior
of $\Omega_k$ with the change of $c$, and figure $4c$ exhibits the
locations of peaks of $\Omega_k$ change with the change of $c$.}
\end{figure}
\begin{figure}[t]
\includegraphics[width=6cm,height=6cm,angle=0]{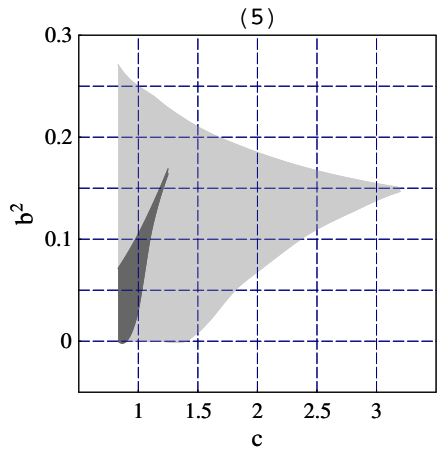}
\caption{This figure shows the constrained parameter space of $b^2$ and $c$.} \label{f1-3}
\end{figure}
\begin{figure}[t]
\includegraphics[width=6.7cm,height=6.58cm,angle=0]{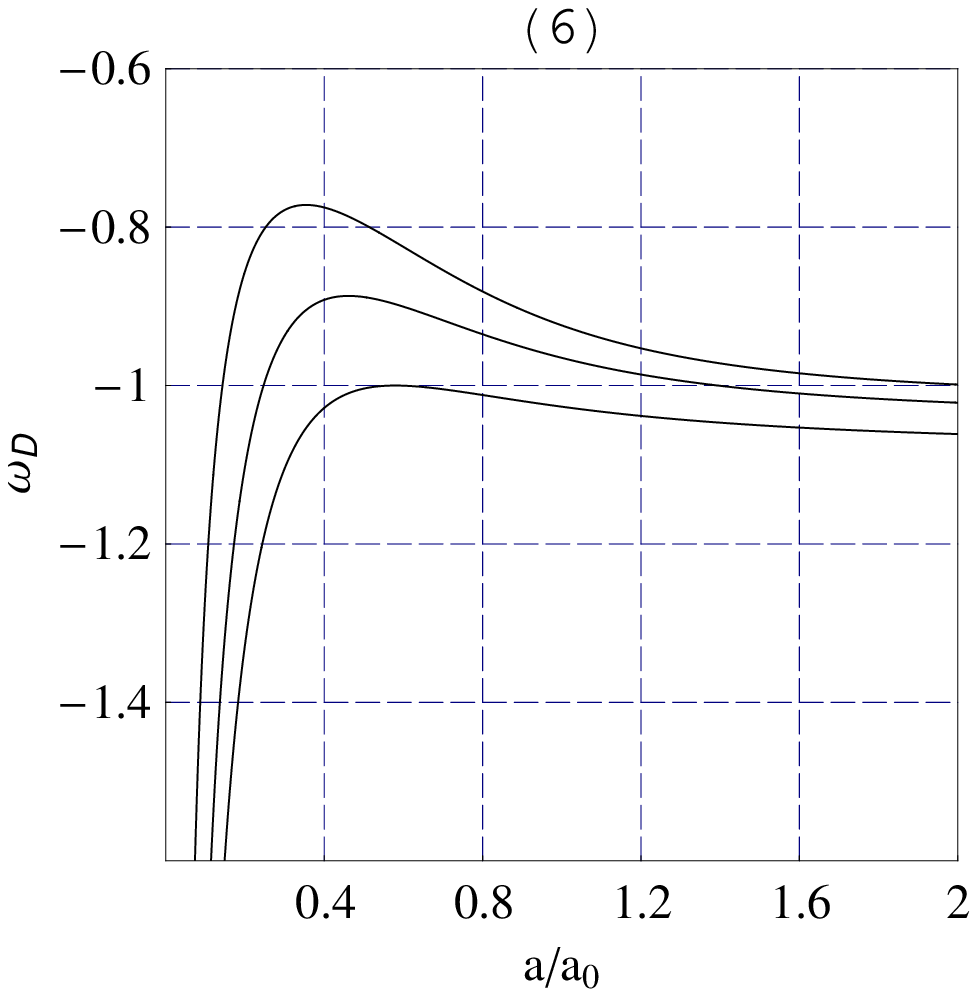}
\caption{Here we show the behavior of $\omega_D$ which crosses the value
-1.} \label{f1-4}
\end{figure}

Considering the Gibbons-Hawking entropy in a closed universe, $S=\pi L^2$,
we require $c^2\geq\Omega_D/(1+\Omega_k)$ to satisfy the second law of
thermodynamics \cite{7}. By choosing the present values, such that
$\Omega_{D0}=0.7, \Omega_{K0}=0.02$, we need $c\geq 0.83$. For fixed small
values of $c$ (but $c\geq 0.83$), we observe that with the increase of
$b^2$, the peaks of $\Omega_k$ appear at a smaller scale of the universe
and peaks increase with the increase of $b^2$. Since
$\Omega_k=1/(aH)^2=1/\dot{a}^2$, the location of $\Omega_{k,max}$
corresponds to the minimum value of $\dot{a}$, which is the starting point
of the acceleration $(\ddot{a}=0)$. To describe the present accelerated
expansion of our universe, the location of the peak of $\Omega_k$ should
appear before the present scale, $a\leq a_0$. This gives the lower bound
on the value of the coupling between DE/DM.

On the other hand, by taking
account of the Friedmann equation $(\dot{a}/a)^2=(1+r)\rho_D/3-1/a^2$, we
have $\Omega_T=1+1/\dot{a}^2=[1-1/f(a)]^{-1}$, where we defined the
function $f(a)=(1+r)H^2\Omega_D a^2=1+1/\Omega_k$. Noting that $\Omega_T$
is close to one since the early universe, when $a\rightarrow 0$, which
indicates that the second term in square brackets in $\Omega_T$ must be
small, we may expand the expression
$\Omega_T\dot{=}1+\Omega_k/(1+\Omega_k)=(\Omega_k+\Omega_T)/\Omega_T$.
This corresponds to requiring that $\Omega_k<<1$ at any time. Using this
``near-flatness" condition and putting by hand that $\Omega_{k,max}\leq
0.04$, we can obtain the allowed $b_{max}^2$ which satisfies the
``near-flatness" condition for fixed small values of $c$.

The characters discussed above are shown in Fig.1. Fig.1$a$ is the
evolution of the DE. Fig.1$b$ shows the behavior of $\Omega_k$. We see
that with larger coupling $b^2$, the DE dominates earlier, so that the
acceleration starts earlier. Fig.1$c$ shows the location of the maximum
value of $\Omega_k$ with the change of $b^2$. It is clear that when
$b^2=0$, $\Omega_{k,max}$ appears at $a/a_0=1.08$ for $c=1.5$, which shows
that to enter the accelerated expansion before the present time, the
interaction between DE and DM is required. The minimum coupling between DE
and DM to drive the universe in the accelerated expansion is $b^2=0.05$
when $c=1.5$.

With the increase of the value of $c$, we observed that allowed range of
$b^2$ by conditions we mentioned above becomes small. For $c>3.2$, we find
that for all values of $b^2$, $\Omega_{k,max}$ appears after the present
scale $a_0$, which is shown in Fig.2. Therefore, in order to accommodate
the acceleration starting before the present time, we found the maximum
value of $c$ as given by $c_{max}=3.2$.

Numerically we have also observed the evolution behaviors for fixed small
coupling between DE and DM with the change of the constant $c$, see
Fig.3. We found that with the increase of $c$, the position of the peaks
of $\Omega_k$ move to appear at a larger scale $a$. Requiring that the
acceleration begins before the present time, we get the allowed $c_{max}$
for the fixed small $b^2$, say $c_{max}=2.4$ for $b^2=0.1$. Combining the
lower bound of $c$ from the second law of thermodynamics, for fixed small
value of $b^2$, we have the parameter space of the constant $c$.

The property described above changes  drastically when
$b^2>b^2_{cr}=0.14$. For fixed $b^2>0.14$, the results are exhibited in
Fig.4. We saw that with the increase of $c$, the peaks of $\Omega_k$
appear for smaller values of $a$ and values of $\Omega_{k,max}$ increases
with the increase of $c$. The permitted $c_{max}$ can be gotten by using
the ``near-flatness" condition, while $c_{min}$ can still be gotten from
the second law of thermodynamics. With the increase of $b^2$, the allowed
parameter space of $c$ becomes smaller. The range of $c$ vanishes when
$b^2$ reaches $b^2_{max}=0.31$.

The allowed parameters' space of $b^2$ and $c$ discussed above satisfying
the ``near-flatness" condition and accommodating the accelerated expansion
of our universe happened before the present era is shown in the yellow
area of Fig.5.

In order to explain the recent observation that DE experiences a
transition from $\omega_D>-1$ to $\omega_D<-1$, we have further
constrained the parameters' space of $b^2$ and $c$. For given $c$, we
found that the DE transition can happen earlier for stronger coupling
$b^2$ between DE and DM, which can be seen from Fig.6. If $c$ is bigger,
the allowed $b_{min}^2$ that accommodates this DE transition increases. On
the other hand, for $c>1.2$, $\omega_D$ will only tend to $-1$ from above
but never crosses it when $a\rightarrow \infty$ for any values of
$b^2$. The parameter space is shown in the red region in Fig.5. If the
future more accurate observations can tell us the exact location of this
DE transition happened, this parameter space can be further
constrained. The value of $c$ around one ($c\in[0.83,1.2)$) is
interesting, since this could serve as a support that the IR regulator
might be simply related to the future event horizon. With the future more
precise data, this question can be answered exactly.

We also have to point out that the interaction between DE and DM may
change some properties of DM clumping. However, this can happen only at
a larger time scale, namely after they "thermalize". The question is
whether there has been time for both to thermalize or not. We think
that they have not! Indeed, the two sectors interact weakly, with an
interaction that contains Hubble constant  therefore the scale of 
thermalization should be very large. Today there has certainly not have 
passed enough time for that. In fact, we know that the expansion is quite 
recent, thus we are actually at the beggining of the DE dominated era.
The interaction cannot change the smooth property of the DE, which is, to
our mind, an observational fact however it will influence the
clumpy behavior of the DM, certainly not yet thermalized. After the 
transition from $w_D>-1$ to $w_D<-1$, DM might presumably become smoother 
than before. This question however has to be focused in the framework of
structure formation.

With the interaction between the DE and DM, neither of them can evolve
separately. The interaction alters the evolution of matter perturbation
and
the formation of cosmological structure. The study of the evolution of
sub-Hubble linear perturbations in the universe with the DE coupled to DM
has been carried out \cite{amendola} and it was found that the
perturbation
grows for $w_D>-1$, while it is always suppressed in the $w_D<-1$ case. We
expect that this result will also hold in our model.

In summary, we have extended our interacting holographic DE model
\cite{11} to the universe with spatial curvature. By imposing the
``near-flatness" condition and requiring that at the present era we are
experiencing the accelerated expansion, we have obtained the parameter
space of the coupling between DE and DM and the constant $c$ from
holography. To accommodate the transition of DE from $\omega_D>-1$ to
$\omega_D<-1$, we have further constrained the parameter space on $b^2$
and $c$. Furthermore we have obtained the results in a closed universe, 
which is a case mildly favored by recent analysis \cite{8b,16}.

\begin{acknowledgments}
This work was partially supported by  NNSF of China, Ministry of Education
of China, Ministry of Science and Technology of China under grant 
NKBRSFG19990754 and Shanghai Education Commission. The work of
C.-Y. L. was supported in part by the National Science Council under Grant
No. NSC- 93-2112-M-259-011. E. Abdalla's work was partially supported by
FAPESP and CNPQ, Brazil. 
\end{acknowledgments}


\end{document}